\documentclass[journal=aaemcq,manuscript=article]{achemso}
\usepackage{graphicx}
\usepackage{float}
\usepackage{chemformula} 
\usepackage{amsmath}
\usepackage{mathtools}
\usepackage{bm}
\usepackage[T1]{fontenc} 
\usepackage{placeins}
\usepackage{multirow}
\usepackage{multicol}
\usepackage{subcaption}

\author{Nasim Tavakoli}
\affiliation[amolf]{Center for Nanophotonics, AMOLF, Science Park 104, NL1098XG, Amsterdam, The Netherlands}

\author{Wonjong Kim}
\affiliation[epfl]{\'Ecole Polytechnique F\'ed\'erale de Lausanne, Switzerland}

\author{Tom Veeken}
\affiliation[amolf]{Center for Nanophotonics, AMOLF, Science Park 104, NL1098XG, Amsterdam, The Netherlands}

\author{Dominique Poorten}
\affiliation[amolf]{Center for Nanophotonics, AMOLF, Science Park 104, NL1098XG, Amsterdam, The Netherlands}

\author{Lucas G\"uniat}
\affiliation[epfl]{\'Ecole Polytechnique F\'ed\'erale de Lausanne, Switzerland}

\author{Alok Rudra}
\affiliation[epfl]{\'Ecole Polytechnique F\'ed\'erale de Lausanne, Switzerland}

\author{Anna Fontcuberta i Morral}
\affiliation[epfl]{\'Ecole Polytechnique F\'ed\'erale de Lausanne, Switzerland}

\author{Esther Alarcon-Llado}
\affiliation[amolf]{Center for Nanophotonics, AMOLF, Science Park 104, NL1098XG, Amsterdam, The Netherlands}
\email{e.alarconllado@amolf.nl}

\title[title]
  {Coloured and Semi-Transparent Nanowire-based Solar Cells for Building Integrated Photovoltaics}


\begin{document}

\begin{abstract}
Semi-transparent photovoltaics (ST-PV) provide smart spatial solutions to integrate solar cells into already-built areas. Here, we study the potential of semiconductor nanowires (NWs) as promising ST-PV. We perform FDTD simulations for different PV materials in a wide range of array geometries, from which we compute PV performance next to perceived appearance. Surprisingly we find an unusual compromise between photocurrent and transmittance as a function of NW diameter that enables NW-based PV to outperform theoretical limits of non-wavelength selective ST-PV. We theoretically and experimentally demonstrate the robustness of NW arrays to different illumination conditions. We provide the origin behind the outperforming NW array geometries, which is crucial for designing NW-based ST-PV systems based on specific needs.

\end{abstract}

\section{Introduction}\label{BI:Intro}

Smart spatial solutions to integrate solar cells into already-built areas are of extreme importance to maintain and increase the pace of energy transition of our time. Building-Integrated Photovoltaics (BIPV), which are photovoltaic systems that are used as part of the buildings' construction, are often the optimal way of integrating renewable energy systems in urban landscapes where undeveloped land is scarce and expensive.\cite{Eiffert2000,Filip2019}. BIPV is still an emergent technology with the key main challenge of how to minimize the trade-off between solar power conversion efficiency and aesthetics, mechanical flexibility and durability. Such trade-off is found more critical for semi-transparent PV (ST-PV) systems, where the best demonstrated power conversion is about half of the thermodynamic limit for their respective transparency \cite{Traverse2017,Almora2021}.

In ST-PV, part of the visible spectrum is intended to transmit through. Depending on its broadband spectral response, ST-PV is considered wavelength-selective or non-wavelength-selective, and as such each type has its own respective fundamental efficiency limit. 
While the first is ideally designed to absorb the ultraviolet (UV) and near-infrared (NIR) part of the sunlight and letting the visible part (or a selected part of it) pass through, the latter offers visible transparency by means of reducing the overall absorption in the cell. 
In general, the selective UV and NIR absorption makes wavelength-selective options most efficient ST-PVs, with a theoretical maximum PV efficiency of 21$\%$ at full transparency compared to the 33$\%$ efficiency limit in opaque cells.\cite{Lunt2012} The difference simply arises from the 19 mA/cm$^2$ of photocurrent given by the visible spectrum. While potentially efficient, wavelength-selective ST-PV are so far attained by external photonic control (with sophisticated fabrication methods) or excitonic absorbers (which either suffer from poor efficiency or stability).\cite{Brus2019} Non-wavelength-selective options, on the other hand, are simple in fabrication (e.g. by reducing the amount of absorber material) at the cost of colour control and the largest compromise in conversion efficiency. This highlights the importance of harvesting NIR photons to maximize efficiency \cite{Rahmany2020,Shi2020,Sun2017}.\\

In this work we exploit built-in nanophotonic concepts in nanostructured PV materials to minimize the negative effects of semi-transparency and colour on efficiency in semi-transparent PV.
We propose free-standing semiconductor nanowire (NW) arrays as a new promising solution as it introduces a new class of ST-PV in between wavelength–selective and non–wavelength–selective. Semiconductor NWs are well-known for their large absorption per unit volume, which has been proven useful in photovoltaic and photoelectrochemical applications.\cite{Anttu2013, Krogstrup2013,Wallentin2013,Park2013,Kelzenberg2010,Fan2009} 

Owing to light coupling into leaky waveguiding modes in vertically standing NWs, the absorption spectrum of NW arrays can be tailored by geometry (i.e. NW diameter and pitch distance) without the need for modifying material composition.\cite{Wang:12,Heiss2014,Mokkapati2015,Wu2012} Such a control over the absorption spectrum in NW arrays has enabled completely new concepts for tandem PV designs\cite{Dorodnyy2015,Tavakoli2019,Chen2016} and it offers a smart way to introduce wavelength–selective semi-transparency in single junction solar cells by reducing the photoactive material volume in a different manner. An additional advantage of NW-based ST-PV is their compatibility with high efficiency PV materials such as III-V semiconductors and Si, which opens up the possibility of minimising the compromise between aesthetics and power conversion efficiency often found in other wavelength-selective PV materials. Thanks to peeling-off techniques, NWs can be removed from their growth substrate to any transparent external support for the realisation of semiconductor PV windows.

Here, we use FDTD simulations to estimate the color, transparency and PV performance of free-standing vertical NW array cells based on three common high-efficiency PV materials, namely GaAs, InP, and Si.
We find counter-intuitive interplay between transparency and PV performance in this new class of ST-PV, which we explain from the dispersion of multiple waveguiding orders with NW diameter. We then fabricate free-standing GaAs NW arrays with nine degrees of transparency and color. We examine the effect of illumination angle on their appearance with wide-field optical microscopy. We find that the appearance is fairly constant up to about an angle of 30$^\circ$, which we explain by the competing excitation of Mie resonances and coupling into waveguiding modes.

\section{Results and discussion}
\subsection{Theoretical considerations on Photocurrent vs Transparency}
\subsubsection{Idealized ST-PV}

\begin{figure}[t!]
	\centering
	\includegraphics[width=0.9\linewidth]{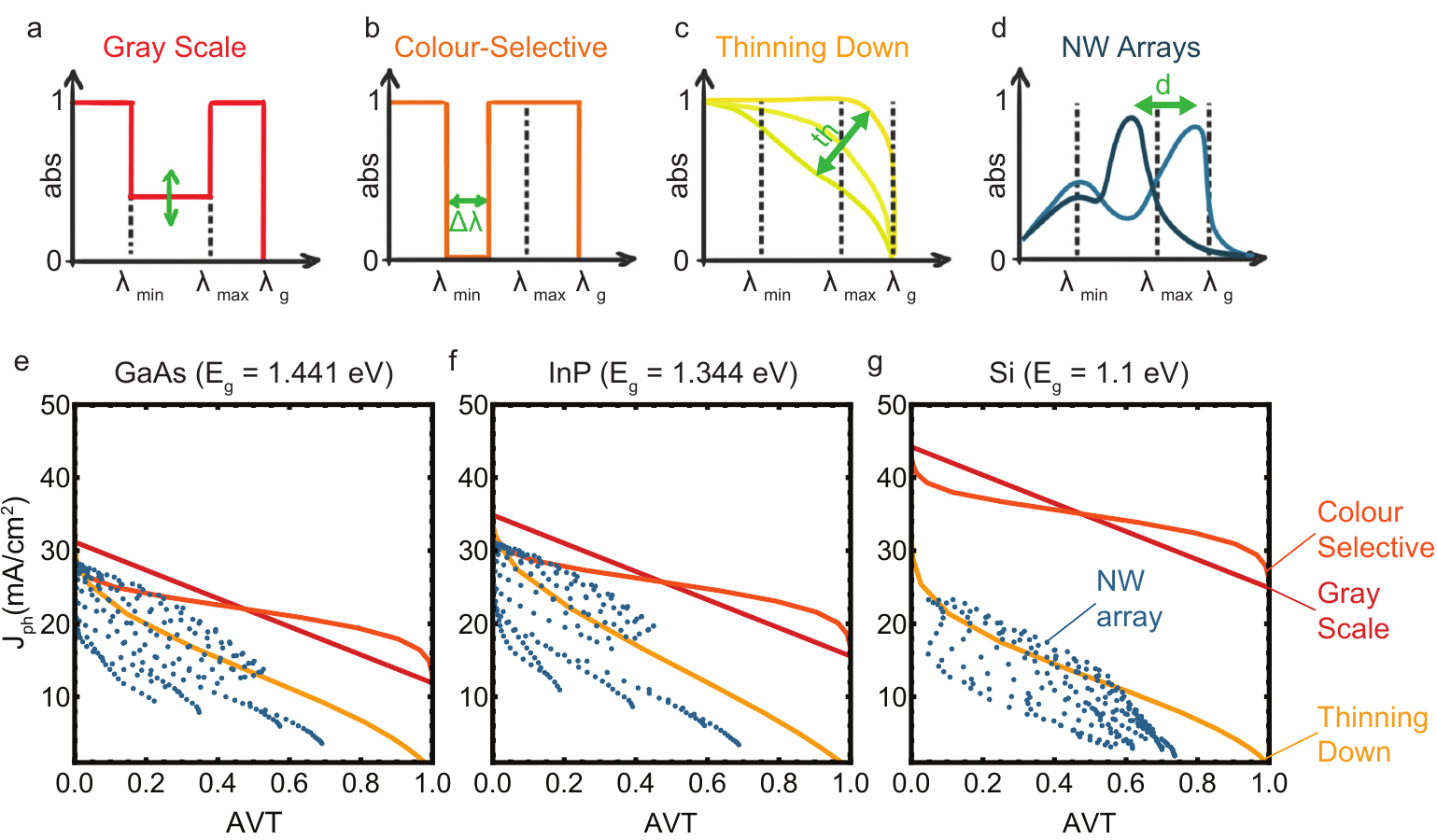}
	\caption{Top: Sketch of ideal broadband absorption spectra for wavelength-selective (a) and b)), non-wavelength-selective (c) and NW array-based (d) ST-PVs. Shades of yellow in (c) illustrate the effect of absorber thicknesses to the absorption spectrum, and shades of blue in (d) indicate the effect of NW diameter. Bottom: Calculated AM1.5G photocurrent density (J$_{ph}$) versus Average Visible Transmittance (AVT) for all ST-PV approaches and for three common PV materials: e) GaAs, f) InP, and g) Si. Solid curves correspond to the idealised (a) to (c) scenarios. Blue data points are given by FDTD simulated absorption in NW arrays with various diameter and pitch distances}
	\label{fig:BI:Schemes}
\end{figure}

First, we illustrate ideal cases of wavelength and non-wavelength selective PV approaches, and their respective maximum PV performance as a function of degree of transparency. The degree of visible transparency is quantified by the Average Visible Transparency, or AVT \cite{Lunt2012,Yang2019}. 
In ideal wavelength-selective cells the absorption spectrum resembles that of a bandpass filter, where the transmission fraction and bandwidth determine the cell’s AVT. Figure \ref{fig:BI:Schemes} a and b showcase two ideal wavelength-selective absorption spectra, where the AVT is given by a constant transmission fraction throughout the whole visible spectrum (labelled as “Gray scale”) or it is given by the bandwidth of a $100\%$ transmissive spectral band (labelled as “Colour selective”). In the latter, we have chosen that the bandpass minimal wavelength is fixed at 390 nm.
While in the “Grey scale” case, the AVT value is directly given by 1-Absorption, to calculate the AVT in the “Color-selective” case one needs to take the cell transmission and photopic response of the human eye into consideration. More details on the AVT calculations and photopic response of the human eye can be found in the SI.\\

Next to the two wavelength-selective ideal show-cases, Figure \ref{fig:BI:Schemes} c shows the absorption spectrum of a non-wavelength-selective approach, where transparency is induced by thinning down the absorber layer (labelled as “Thinning down”). This is the simplest way to achieve transparency in a cell from the fabrication point of view. In this case, one can consider the Beer-Lambert law to obtain the thickness-dependent absorption for a given absorber material, sketched in shades of yellow in Figure \ref{fig:BI:Schemes} c.
Figure \ref{fig:BI:Schemes} e, f and g show the maximum photocurrent as a function of AVT for three common high-efficiency PV materials (namely GaAs, InP, and Si), where we assume ideal absorption in the three semi-transparency scenarios described above. In all three ideal semi-transparency scenarios (\textit{grey scale}, \textit{colour selective} and \textit{thinning down}), we have neglected reflection at the glass or solar cell, thus representing the respective upper limit performance. 
As expected, the more transparent the solar cell the lower the photocurrent. Note that the \textit{thinning down} approach is the least performing semi-transparency solution with total loss of photocurrent at 100\% transparency. Contrarily, in the two wavelength-selective approaches (\textit{grey scale} and \textit{colour scale}) the photocurrent loss from AVT$=0$ to AVT$= 1$ is $19$ mA/cm$^2$, regardless of material. 
It is also interesting to note, that between the two wavelength-selective approaches, \textit{gray scale} seems better suited for ST-PV with AVT$<50$\%, while for larger transparencies, \textit{color selective} offers higher performance for the same transparency.

\subsubsection{NW-based ST-PV}

\begin{figure}[t!]
	\centering
	\includegraphics[width=0.4\linewidth]{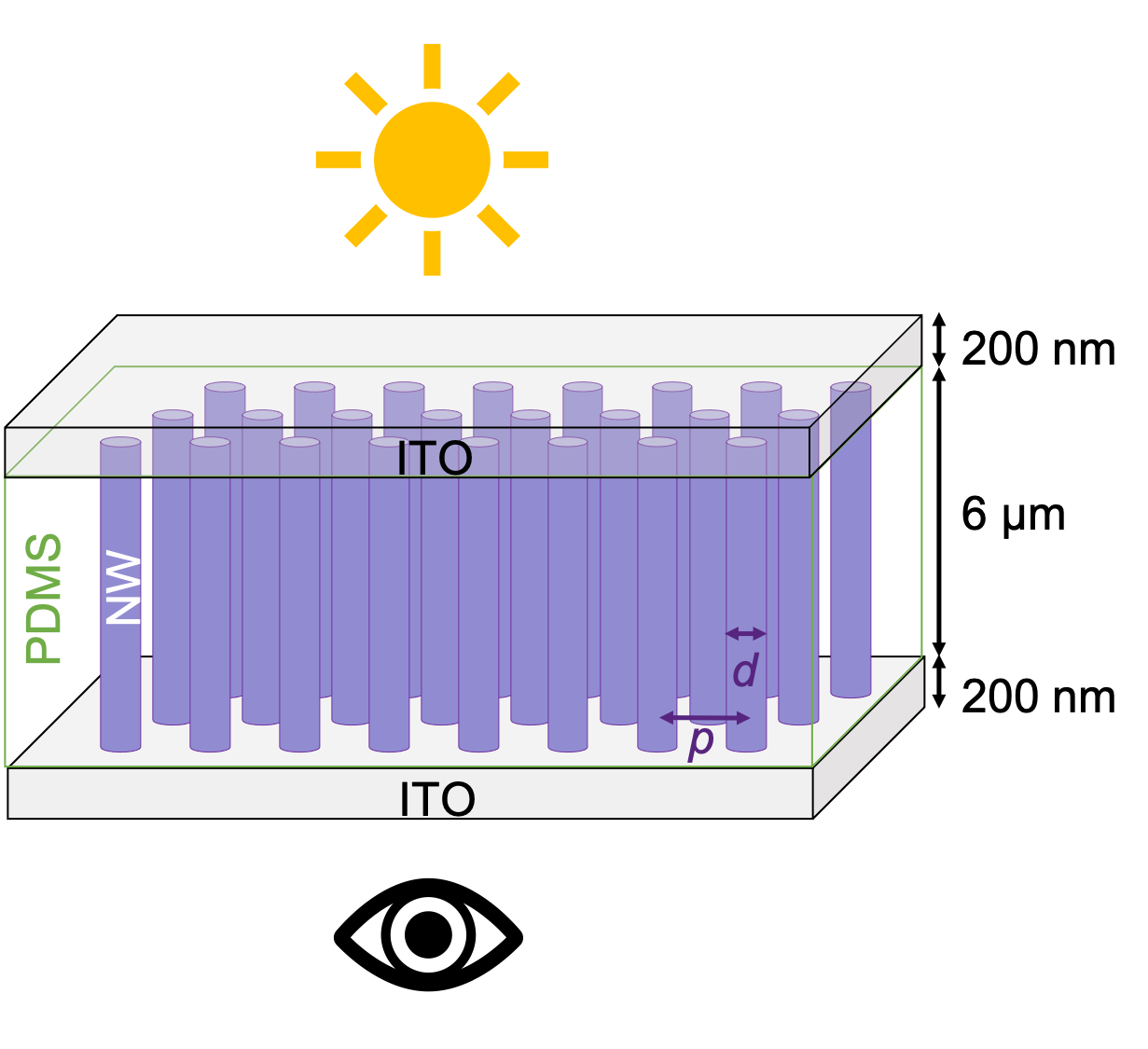}
	\caption{Sketch representation of the NW-based semi-transparent solar cell configuration. A square array of vertically standing NWs are embedded in a polymer matrix and sandwiched between two ITO layers. The thickness of all layers is fixed to the values indicated in the drawing, while the diameter and pitch distance are tuned to tailor different appearance in transmission.}
	\label{fig:BI:ColouredWindows}
\end{figure} 

We now introduce the free-standing NW-based PV stack as a ST-PV solution. Square NW arrays of either GaAs, InP or Si are embedded in a PDMS layer that provides mechanical stability which is sandwiched between two 100-thick ITO layers as transparent electrode contacts (see sketch in Figure \ref{fig:BI:ColouredWindows}). For the PV effect one requires asymmetric contacts, which may be achieved via doping within the NW\cite{Wallentin2013,Otnes2018,VanDam2016,Aberg2016} or by the introduction of thin carrier blocking layers or selective contacts\cite{Espinet-Gonzalez2019,Raj2019,Raj2019-review}. Additional passivation layers at the large surface area of the NWs is also crucial to maximize the open circuit potential.\cite{Wallentin2013,Otnes2018,Espinet-Gonzalez2019} Since the optical response of the arrays is not likely strongly affected by such thin layers or doping profiles\cite{Kelzenberg2010,Zhong2016}, we thus consider the simplified case of two symmetric ITO contact layers. We have examined periodic arrays of NWs with diameters between $50$ and $200$ nm, and pitch distances from $300$ to $1000$ nm. The NW length was fixed to 6 $\mu $m for all materials, since this length is a fair compromise between absorption efficiency and fabrication feasibility. More details of the simulation set-up can be found in the SI. \\

Using the transmission and absorption spectra as obtained from FDTD simulations of the NW-based PV stacks we calculate the photocurrent and AVT for a series of NW array geometries (blue data points in Figures \ref{fig:BI:Schemes} e, f and g). Each data point represents a single array with particular combination of NW diameter and pitch distance.
From Figure \ref{fig:BI:Schemes} e, f and g, it is clear that NW array ST-PV offers a wide range of possible AVT values, up to about 70$\%$ transparency which is a similar value to what has been demonstrated so far with other material systems.\cite{Almora2021} For reference, AVT values for residential windows varies between 15$\%$ and 90$\%$ depending on how tinted or clear the windows are and as a general rule of thumb, glass with AVT above 60$\%$ looks already clear\cite{Fisette2003}. Such a broad range of AVT possibilities for the same single-material in NW arrays is a first indicator of their potential for ST solar cell designs for different implementation areas; ranging from dark, coloured to tinted clear ST glazing.

From the PV performance perspective, some NW array geometries show higher photocurrent values than those given by the \textit{thinning down} approach for the same AVT. Furthermore, some NW array geometries even outperform the \textit{color-selective} approach at low AVTs ($< 20\%$ for GaAs and $< 18\%$ for InP). This result is very impressive given the fact that reflection and parasitic absorption are actually considered in the NW array PV case, contrary to the other idealised scenarios. We address the origin of such a broad performance span in the following section.

\subsubsection{Effect of NW geometry on performance and transparency}

\begin{figure}[t!]
	\centering
	\includegraphics[width=0.8\linewidth]{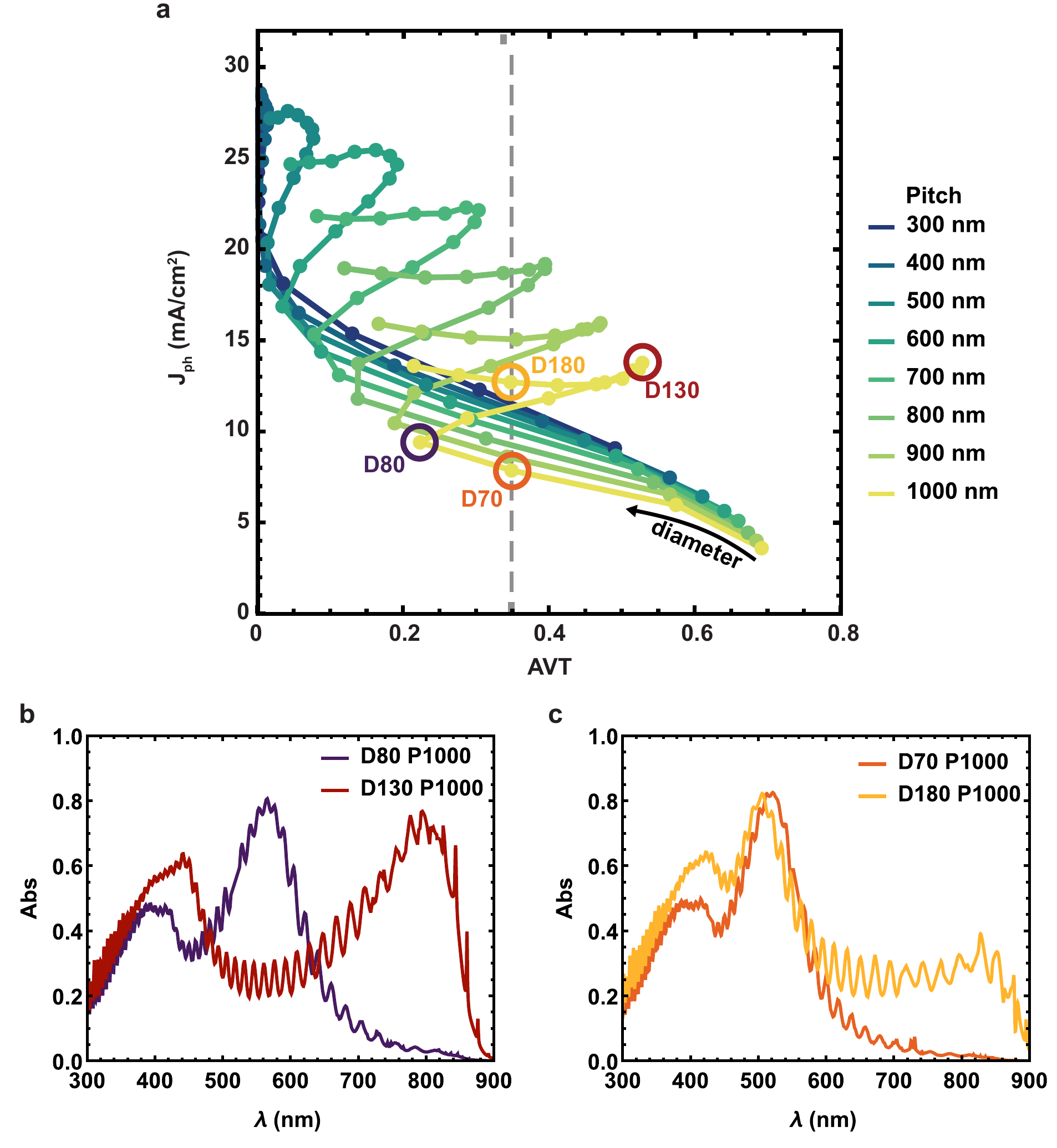}
	\caption{a) Simulated J$_{ph}$ vs. AVT of representative GaAs NW arrays from Figure \ref{fig:BI:Schemes}e collected by pitch values as indicated by the legend.
	b) and c) Absorption spectra for data points highlighted with circles in (a).
	}
	\label{fig:BI:JscAVT}
\end{figure}

To get a better understanding of the mechanism behind the broad span of transparency and photocurrent values in the NW arrays, the same set of data is now plot by colour coding the pitch distance. Figure \ref{fig:BI:JscAVT}a shows the change in AVT and photocurrent with increasing GaAs NW diameter (data points from right to left) for eight different pitch distances, color-coded as indicated by the legend. Similar graphs are obtained for the other two PV materials, shown in the SI.  For each pitch distance, the photocurrent vs AVT displays a sort of z-shape. This effect is most evident for the largest pitch distances. For the smallest diameter, the AVT shows the highest values in each of the pitch series. As one may expect, as the diameter is increased the AVT decreases at the same time as the photocurrent increases. However, we find a surprising switch in trend for diameters bigger than 80 nm. For diameters larger than 80 nm and up to 130 nm, the photocurrent increases along with transparency. Again, at a diameter of 130 nm, the trend switches once more, leading to the more standard anti-correlated photocurrent vs AVT. The unprecedented correlation between photocurrent and AVT that we observe for a certain range of NW diameters is an excellent new opportunity to achieve ST-PV based on absorber reduction concepts with efficiencies beyond those obtained from thinning down the absorber. In the following, we explain this phenomenon by the diameter-dependent waveguiding properties of vertical semiconductor NWs.

First, we take a closer look at the absorption spectra (Figure \ref{fig:BI:JscAVT}b) of the two GaAs NW arrays with $1 \mu$m pitch distance where the AVT vs photocurrent changes trend, namely for 80 nm and 130 nm in diameter (labelled as D80 and D130, respectively). Each geometry result in the lowest/highest transparency for similar photocurrent values among their neighbouring data points, respectively. The peaks in absorption found at $\sim 550$ nm for D80 ($\sim$ 450 and 800 nm for D130), are well known to arise from the efficient light coupling to HE01 (HE01 and HE02) waveguiding modes of the nanowires \cite{Heiss2014,Frederiksen2017}. In the case of 80 nm GaAs NWs embedded in PDMS, the first order waveguiding mode sits right at the peak of the human eye photopic response, thereby becoming the least transparent array of all the diameter series. Increasing the NW's diameter causes a red-shift in the absorption peak(s). Because of the diameter-induced red-shift of the absorption peak, the transparency increases with diameter. At the same time, a new absorption peak appears at shorter wavelengths due to higher order waveguiding mode. Consequently, the photocurrent also increases with diameter.

The dual increase in AVT and photocurrent occurs until the nanowires reach the diameter of 130 nm, at which point the first order waveguiding mode coincides with the GaAs bandgap ($\sim$850 nm). As the diameter further increases, the efficient coupling to the first order waveguiding mode no longer sits within the absorption spectrum of GaAs and thus the photocurrent sharply decreases. Notice that this effect is less evident in Si due to its indirect bandgap nature (see SI). In parallel to the photocurrent decline, the higher order absorption peak at shorter wavelengths shifts towards the visible spectral range as the diameter increases, which in turn reduces the AVT. As a result, the anti-correlation between AVT and photocurrent is attained again. We observe that at smaller pitch distances the z-shape in the photocurrent vs AVT is softer, which we attribute to the optical cross-talk between NWs broaden the absorption peaks.

An interesting consequence of the z-shape behaviour is that more than one photocurrent value can be achieved for the same AVT value. See for instance the two datapoints D70 and D180 marked in Figure \ref{fig:BI:JscAVT}a. While both array geometries show an AVT value of $\sim 0.35$, the photocurrent is 1.6 times larger for the larger diameter (i.e. 7.8 and 12.7 mA/cm$^2$). The absorption spectra for these two array configurations as shown in Figure \ref{fig:BI:JscAVT}c reveal that both arrays display very similar absorption peaks although arising from different order waveguiding modes. 
However, for the large diameter case there is the tail absorption related to the first order waveguding peak in the NIR (around 850 nm), which is the responsible for the larger J$_{ph}$ value in D180.\\

In summary, the sudden correlated increase in AVT and photocurrent arises from the fact that there are multiple absorption peaks in the spectra of vertically standing NWs. For certain NW geometries, the two peaks are just positioned at each side of the visible spectral range, resembling the absorption spectra for wavelength-selective solar cells. The nanowire array approach to make semi-transparent coloured solar cells is a very powerful concept: by tuning their geometry through diameter and pitch, the interplay between AVT and J$_{ph}$ can be adapted to maximize performance without the need to fine tune the material composition.

\subsubsection{Effect of NW geometry on cell appearance}

\begin{figure}[t!]
	\centering
	\includegraphics[width=0.9\linewidth]{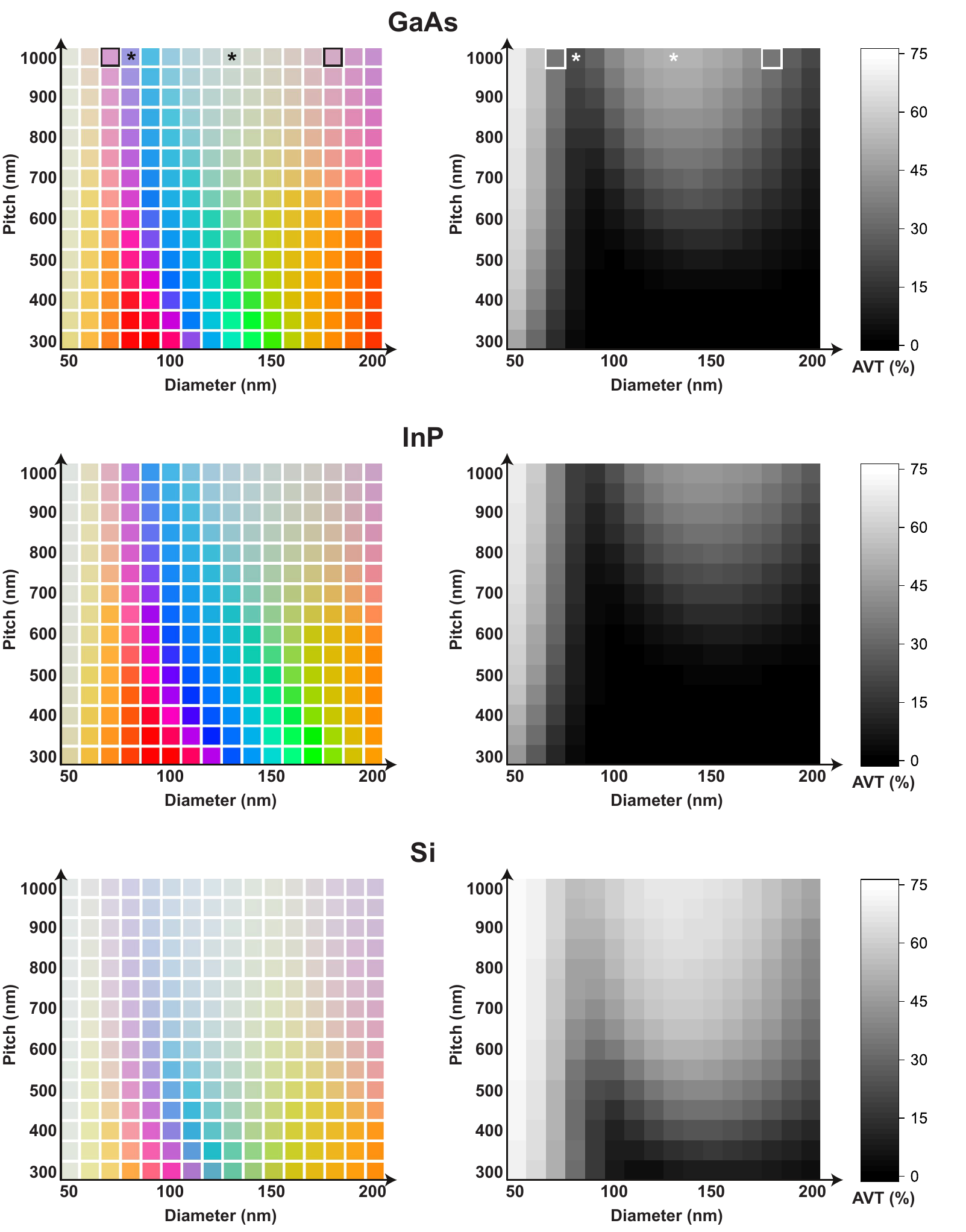}
	\caption{(left) Calculated perceived colour for NW array cells based on GaAs, InP, and Si (top to bottom, respectively). (right) AVT matrix representation of the same NW array cells. The squares marked with an asterisk correspond to the color and AVT for geometries whose spectra are shown in Figure \ref{fig:BI:JscAVT}b. At these two diameter values, the arrays display the lowest and highest AVT for a given pitch, respectively. 
}
	\label{fig:BI:ColourPalette}
\end{figure}

Next, we focus on the NW array solar cells daylight appearance for the same geometries considered in the previous section. Figure \ref{fig:BI:ColourPalette} a and b show both the AVT and simulated appearance in transmission, respectively, of GaAs, InP and Si NW arrays in a matrix form. For the appearance, we use the FDTD simulated transmittance, the D65 daylight spectrum and the Colour Matching Functions (CMF) from CIE 1964 standard \cite{CMF}. Further details on the color calculation and representation are found in the SI. 
At first glance, Figure \ref{fig:BI:ColourPalette} proves that NW arrays open up the possibility of creating a beautiful and diverse colour palette with a wide range of transparency for integrating solar cells into buildings. Note that the color squares represented in \ref{fig:BI:ColourPalette}a contain information on both color hue as well as transparency. 

By comparing the AVT matrices with those for the colors in Figure \ref{fig:BI:ColourPalette}, one can notice that high transparency results into less saturated colors and viceversa. Colours are highly saturated with AVTs up to 0.4 for arrays with small pitch distances. By contrast, arrays with pitch distances larger than 600 nm show less saturated colours arising from the larger degree of semi-transparency (AVT between 0.4 to 0.7). While it is sensible that the decreased NW density from increasing pitch distance induces a larger degree of transparency, we observe that the pitch also has an effect on the hue, particularly for pitch distances shorter than 600 nm (or $<$500 nm in the case of Si). This is observed by the gradual shift of the colour pallet to larger diameters as the pitch decreases.
The hue is also highly dependent on NW diameter. At a given pitch distance, the color follows the inverse of the rainbow order by increasing diameter and eventually repeating colour at larger diameters (see for instance the two squares highlighted in the GaAs colour matrix). Such a trend of color hue with diameter is not unexpected given the red-shift of waveguiding-induced absorption peaks with increasing diameter, as explained in the previous section.

\begin{figure}[t!]
	\centering
	\includegraphics[width=0.5\linewidth]{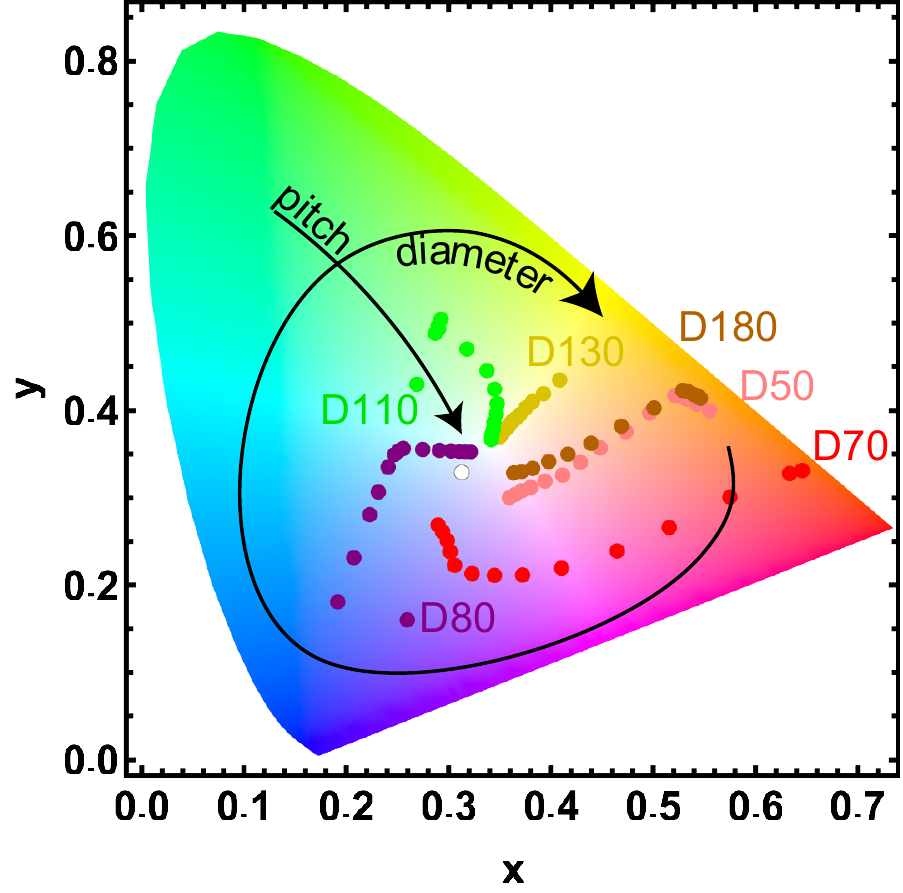}
	\caption{Chromaticity diagram overlapped with chromaticity values for GaAs NW arrays with diameters 50, 70, 80, 110, 130 and 180 nm and pitch distances from 300 to 1000 nm. The increase in diameter and pitch is indicated by black arrows.
	}
	\label{fig:BI:Chromaticity}
\end{figure}

A different and more objective way to describe the effect of NW array geometry on the appearance is by using the chromaticity diagram (Figure \ref{fig:BI:Chromaticity}). The chromaticity x and y coordinates are given by the normalized tristimuli, as described in the SI. More intuitively, in the chromaticity diagram the hue is indicated by the angular component from the central white point, while the radial component indicates colour purity (i.e. pure colours are represented at the edge of the diagram). Overlapped with the chromaticity diagram in Figure \ref{fig:BI:Chromaticity}, we include datapoints indicating the colour evolution of GaAs NW arrays with changing pitch distance for seven different NW diameters (chromaticity plots for the other two material systems can be found in the SI). Arrays with same NW diameter are represented by datapoints of the same colour (e.g. red datapoints correspond to arrays of 70 nm in diameter). As expected from the colour pallets, we find that the various array geometries cover most of the chromaticity plot. Looking at the farthermost datapoint from the center of each diameter configuration, it is clear that the hue changes with diameter in a cyclic fashion from red to blue, green, yellow to red again (represented by the thick black arrow). This repeat in colour arises from the fact the higher order waveguiding mode appears at the same spectral position as the diameter is increased. 

With increasing pitch distance (as indicated by the thin arrow), the array appearance is brought towards the central white point, arising from the increase in transparency. However, it is interesting to note that while there is a fairly straight radial shift in the pitch-induced appearance change for the diameters of 130, 140 and 200 nm, there is also a pitch-induced hue shift in the case of diameters of 80, 90 and 110 nm. This pitch-related hue change is not just an effect that occurs due to the small diameters (as it does not occur for smaller diameters like 70 nm), but rather associated to the spectral position of the waveguide-induced absorption peak for a given NW diameter. 

\subsubsection{Effect of illumination geometry on appearance}

\begin{figure}[t!]
	\centering
	\includegraphics[width=0.9\linewidth]{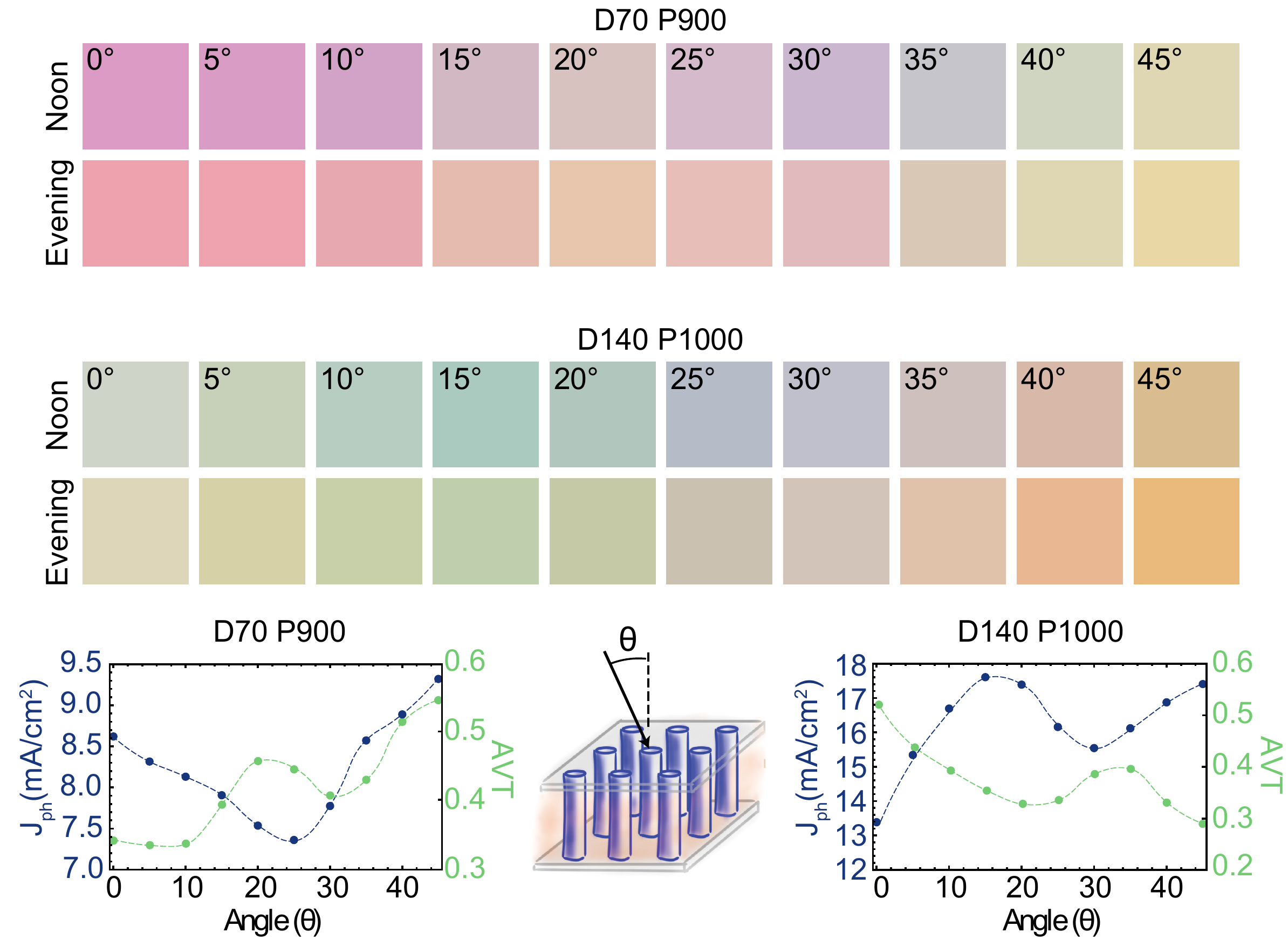}
	\caption{(Top) Perceived colour in transmission for two geometries of GaAs NW arrays (diameter 70 and pitch 900 marked as D70 P900, and diameter 140 nm and pitch 1000 nm marked as D140 P1000) under different angle of incidence, $\theta$ (shown in the sketch). In both cases, the perceived colour is calculated for both the standard daylight spectrum at noon (D65) and the standard evening spectrum. (Bottom) J$_{ph}$ and AVT as a function of $\theta$.}
	\label{fig:BI:AngledFDTD}
\end{figure}

One key application of semi-transparent PV is to be used as window components. Given that the illumination geometry changes through the day/year, it is important to address the influence of angle of incident sunlight on the appearance and the PV performance of ST-PV. To evaluate this effect on the NW solar cells, we have performed FDTD simulations under different angles of incidence ($\theta$) of an incoming plane wave onto a GaAs nanowire array stack  (see the sketch in Figure \ref{fig:BI:AngledFDTD}). We have considered two GaAs NW array configurations as study case, which show very different appearance under standard illumination conditions (i.e. daylight at normal incidence): Array (A) corresponds to nanowires of 70 nm in diameter and 900 nm apart; and array (B) composed by nanowires of 140 nm in diameter and 1 $\mu$m apart. Under standard illumination, array A shows a strong pink appearance (AVT $\approx 34\%$), while array B displays as light green with relatively large transparency (AVT $\approx 53\%$).\\

Apart from changes in the angle of incidence, variations in the relative spectral power distribution of daylight are known to occur, particularly in the ultraviolet spectral region, as a function of season, time of day, and geographic location. Thus, besides the standard noon spectrum (D65) we have also considered the standard evening spectrum (based on black body radiation at 4000K) and represented the perceived colors in Figures \ref{fig:BI:AngledFDTD} as a function of angle, under the \textit{Noon} or \textit{Evening} labels, respectively. 
Interestingly, the appearance of both arrays is relatively similar between noon and evening illumination when the angle of incidence is small ($<20^\circ$-30$^\circ$) despite the stronger red component in the evening spectrum (see SI). At those small angles the colour hue is fairly stable in both arrays, however for larger angles the change in colour is more abrupt. At large angles and noon illumination, the color hue clearly shifts from green to blue and orange in array A, and from pink to blue to green/yellow in array B.
Such a spectral change with angle may be expected by the allowed resonant excitation of other modes in the NWs. Earlier works have shown that the absorption in NWs is mainly governed by coupling to waveguiding modes up to angles of incidence around 20$^\circ$ after which Mie resonances dominate  \cite{Grzela2014}. \\

Based on the simulated spectra and the daylight illumination conditions (AM1.5G), we compute the photocurrent and AVT as a function of angle of incidence on the two NW array cells (bottom panels in Figure \ref{fig:BI:AngledFDTD}). It is interesting to see that the transparency of the two array configurations exhibits completely opposite trends with angle of incidence. While the AVT of array D70 P900 steadily increases with angle, the opposite occurs for array D140 P1000. A decrease in transparency and increase in photocurrent with increasing the illumination angle may be expected, as the NW array appears more dense. In fact, a 0.5\% increase in conversion efficiency for an illumination angle of 15 degrees for a GaAs vertical NW array has been measured before. \cite{Ghahfarokhi2016} However, the AVT behaviour observed in array D70 P900 is different. Comparing the absorption spectra versus the incidence angle for these two geometries suggests that the position of the main absorption peak at normal incidence determines whether or not the non-zero angles would be beneficial for the PV performance (see the corresponding absorption spectra at normal incidence in the SI). For the array D70 P900 the peak in absorption at $\theta =0$ is positioned inside the visible range. For larger $\theta$ however, contributions from Mie modes changes the position of the absorption peaks towards the shorter wavelengths, hence outside of the visible range, which results in higher AVT.

\subsection{Experimental demonstration}\label{BI:Experiments}
\begin{figure}[t]
	\centering
	\includegraphics[width=0.7\linewidth]{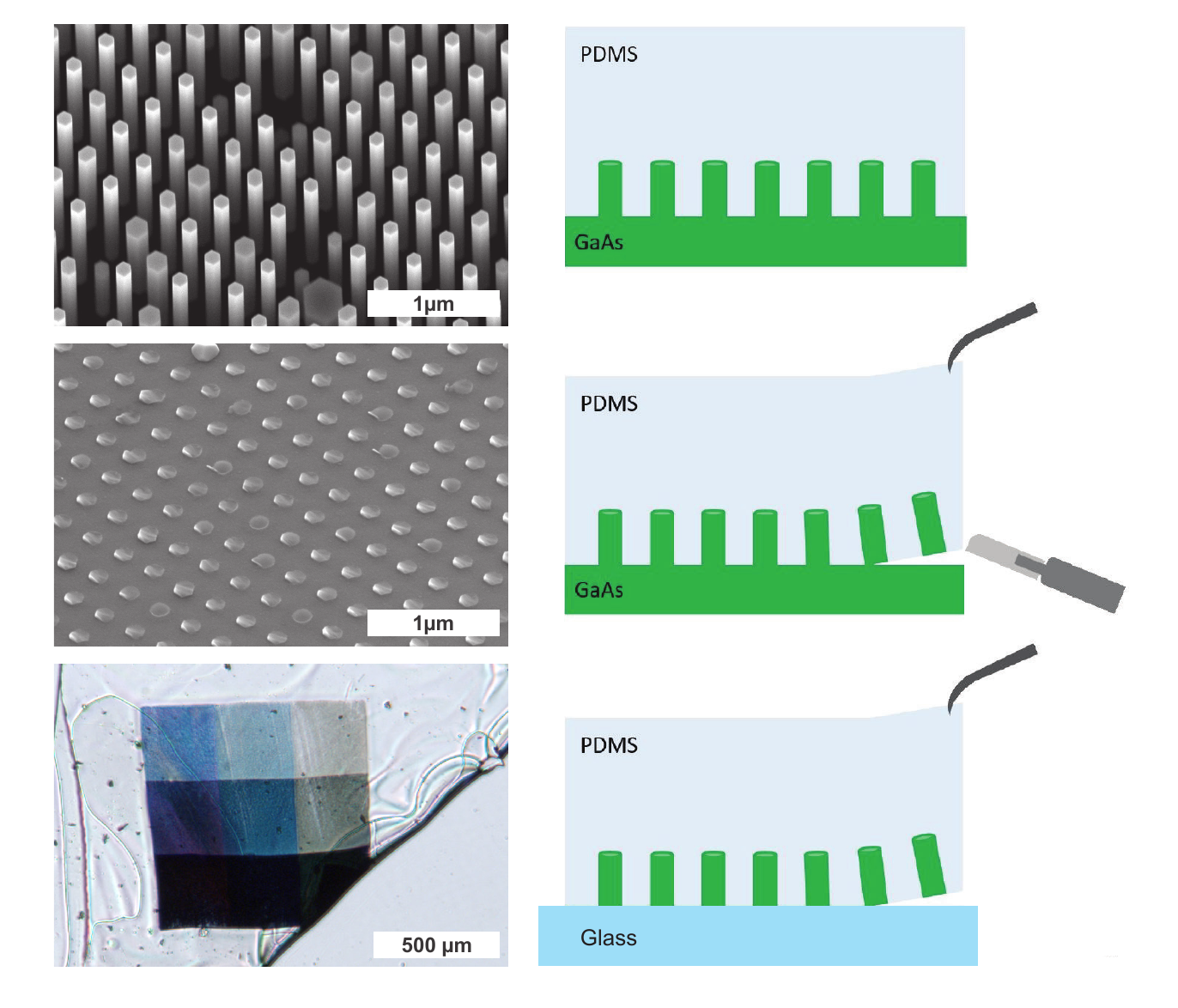}
	\caption{(Right, top to bottom) Sketch showing the steps to prepare NW-polymer composite: embedding the GaAs NW array on GaAs substrate in polymer, cutting and peeling off the NW-polymer composite off the substrate, and placing it on transparent glass holder. (Left, top to bottom) SEM image of the as-grown GaAs NW array, and of the substrate after peeling off. Optical image of the nine fields of PDMS-embedded NW arrays transferred to a clean cover glass}
	\label{fig:BI:PeelOff}
\end{figure}

We now move to the fabrication and characterisation of vertical GaAs NW arrays with different diameters and pitch distances as proof of concept. The arrays were grown by metal organic chemical vapor deposition via selective area epitaxy on a pre-patterned GaAs substrate with periodic arrays of holes (hole diameters of 50, 70, 100 nm and pitch distance of 300, 600, 900 nm). The as-grown sample consists of $3\times 3$ fields of 300×300 $\mu$m$^2$ NW arrays, each with different NW diameter and pitch distance. Owing to lateral overgrowth, the actual NW diameter is larger than that of the openings. We find that the NW diameters are $\sim$100, 145 and 175 nm. 
Owing to the same growth time in arrays with different NW diameter, the NW length is not the same in all arrays, ranging from ~4.5 to ~7 $\mu m$. Further details on the substrate preparation and NW growth can be found in the SI.

As shown schematically in figure \ref{fig:BI:PeelOff}, self-standing arrays were attained by embedding the NW arrays in a PDMS polymer film and subsequently peeling the composite off from the GaAs substrate. A surgical knife was used to cut the polymer area around the NW fields and then undercut the NWs from their base. The NW-polymer composite was removed with the help of tweezers and was placed on a clean cover glass (more details in the SI). Figure \ref{fig:BI:PeelOff} shows the SEM images before and after peeled off, where it is clear that the NWs were nicely cleaved near the NW/substrate interface. The optical microscope image of the PDMS-embedded arrays transferred to a glass substrate is also shown, where NW diameter increases from left to right, and pitch distance from bottom to top. From the optical image, one can already discern different colors (from black to purple, blue and  green) and degrees of transparency (from what appears like satin glass to fully opaque) offered by the NW arrays in such a small span of geometries.

\subsubsection{Optical properties}

\begin{figure}[t!]
	\centering
	\includegraphics[width=0.9\linewidth]{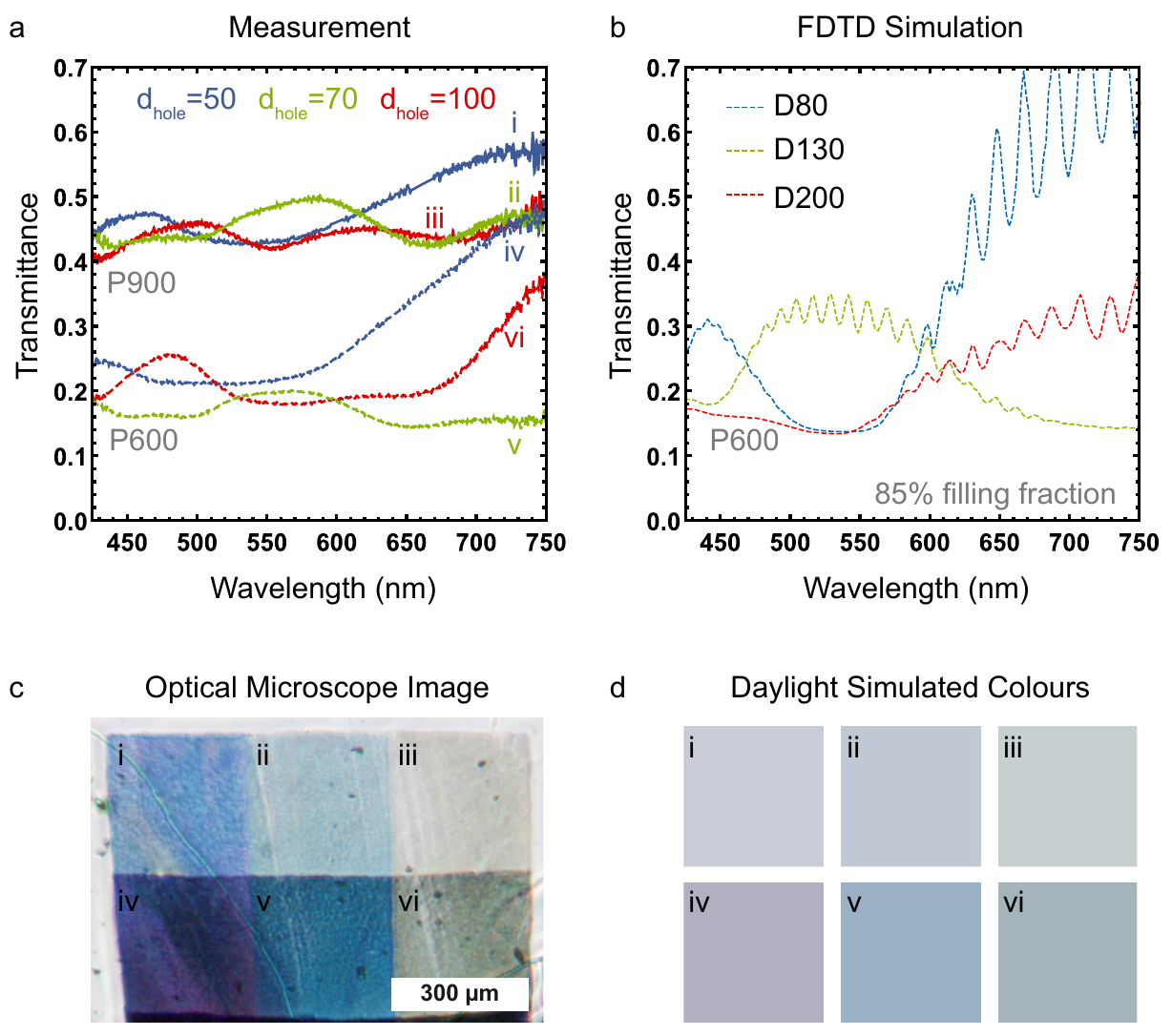}
	\caption{a) Measured transmittance spectra of six NW array fields with nominal hole diameters of 50, 70, and 100 nm (shown in blue, green and red, respectively), and pitch distances of 600 and 900 nm (shown in solid and dotted lines, respectively).  b) FDTD simulated transmittance (1-Absorption-Reflectance) of NW arrays with pitch distance of 600 nm and diameters 80, 130 and 200 nm. The NW absorption has been reduced to 85\% to account for the growth/peel-off yield c) Optical image of the six fields: i, ii and iii are with pitch 900 nm, and hole diameters of 50, 70 and 100 nm, respectively. iv, v and vi have pitch of 600 nm and same sequence in hole diameter. The AVT values for the arrays are i) 44.5\%, ii) 43.7\%, iii) 47.4\%, iv) 23.2\%, v) 19.2\%, and vi) 18.5\%. d) Perceived daylight colour in transmission as calculated from data in (a)}
	\label{fig:BI:witec}
\end{figure}

The optical properties of the NW/PDMS are assessed with optical microscopy in transmission mode. For spectroscopy measurements, a confocal microscope is coupled to a spectrometer.
The transmission spectra and optical image for six of the nine fields are plotted in figure \ref{fig:BI:witec}a and c, respectively. In Figure \ref{fig:BI:witec}c, the top row of arrays (labeled i, ii and iii with increasing NW diameter) corresponds to those with the largest pitch distance (pitch = 900 nm), and thus are more transparent (average AVT is around $20\%$). Oppositely, the bottom row of arrays (labeled as iv, v and vi with increasing diameter) are more dense and thus show more vibrant colors as expected from the theory shown in the previous section. The three arrays with the smallest pitch distance were completely opaque and are thus not discussed here. The blue, green and red curves in Figure \ref{fig:BI:witec}a correspond to geometries with hole diameters of 50, 70 and 100 nm, respectively. Solid (dashed) lines shows the spectrum for the array with pitch distance of 900 nm (600 nm). On average the transmission for the arrays with the largest pitch distance is the highest (with an AVT around $44\%$). 

The three different NW diameters in the arrays give rise to characteristic spectral features in transmission that are not strongly perturbed by the pitch distance. Figure \ref{fig:BI:witec} b shows the simulated transmission spectra for three different NW arrays with diameters of 80, 130 and 200 nm, which are close to the average NW diameter found in the as-grown arrays. For clarity, we only show simulations for one pitch distance (pitch = 600 nm). Here, the transmission has been obtained by considering only 85$\%$ of the NW-PDMS absorption of a perfect array, as to account for defects and missing NWs in the peeled-off arrays. The measured spectra for the two arrays with smallest diameters (blue and green curves) are in good qualitative agreement with the simulated ones, with some discrepancies in the absolute strength of the transmission peaks. We attribute these differences to the different NW length, possible parasitic absorption in the PDMS layer and to the fact that the collection of transmitted light is limited by the objective's NA$=0.8$ in our experiment.  

By contrast, the spectrum of the array with the largest NW diameter (red curves) also qualitatively disagrees with the simulated curve. We have not found a uniform array in our simulations (with pitch distance of 600 nm and 6 $\mu$m in NW length) that presents a similar transmission spectrum to the measured one.  We believe that the differences here are due to the broader distribution of NW sizes and the much shorter NW lengths for the large hole size. 
Based on the measured spectra, we have also computed the expected array appearance under daylight and is represented in Figure \ref{fig:BI:witec}d. It is interesting to note that the colors are slightly different than those in the optical image, as the microscope lamp spectral distribution differs from that of the Sun. Despite subtle differences and smaller contrast, the daylight simulated colors clearly show a progressive color change, from purple to green-blue, with increasing diameter. 

\subsubsection{Appearance under angled incidence}
\begin{figure}[t!]
	\centering
	\includegraphics[width=8 cm]{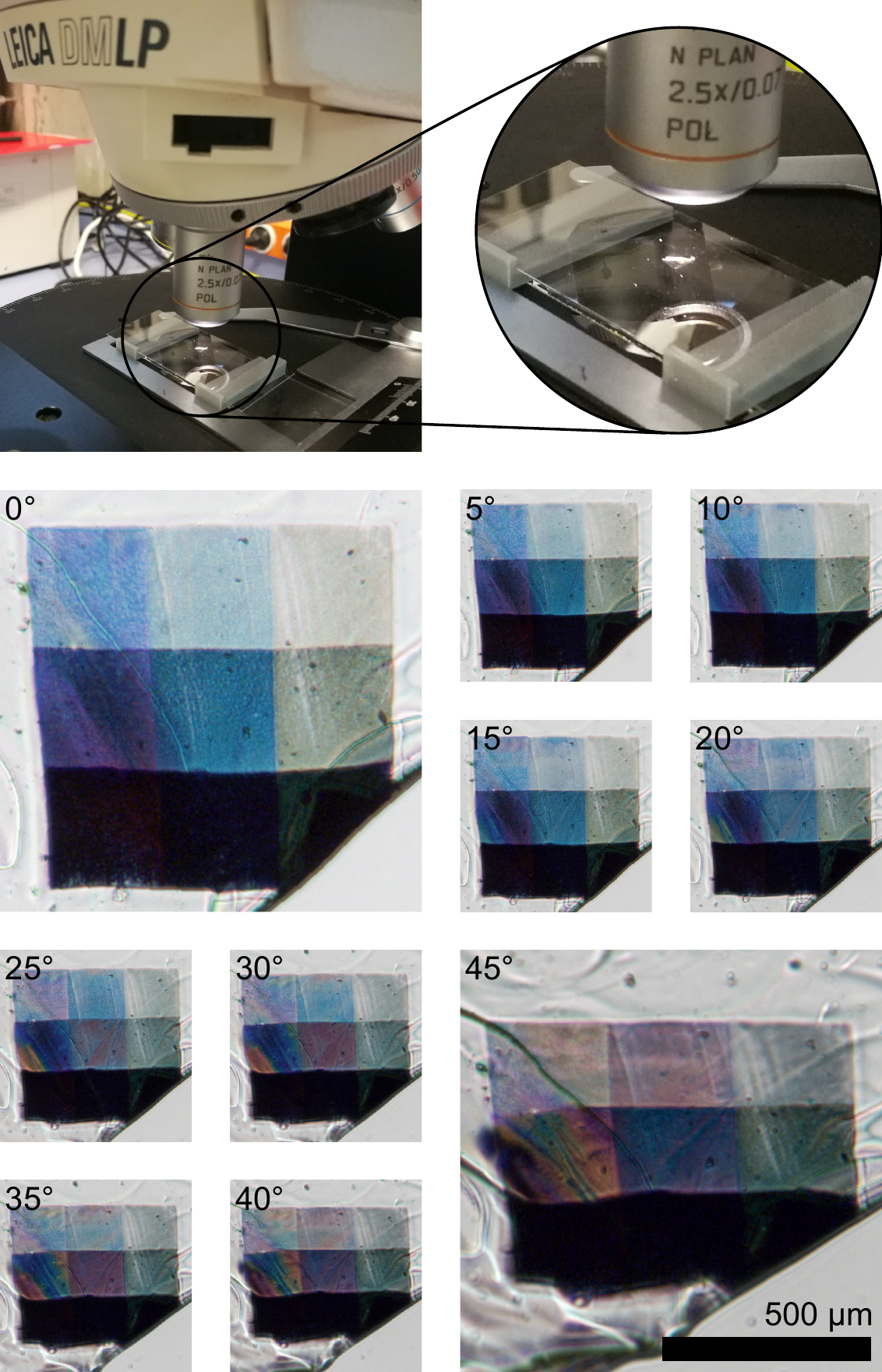}
	\caption{(Top) Picture of the microscope set-up used for the optical imaging of the arrays, and a close-up of the wedged sample stage for the angled incidence measurements. (Bottom) Optical microscopy images of the nine NW array fields transferred to a cover glass, taken under incidence angles from 0 to 45$^\circ$}
	\label{fig:BI:leica}
\end{figure}

We finally also experimentally show the effect of the angle of incidence on appearance of the arrays. To capture images at different angles of incidence, we use an optical microscopy with tailored 3D printed wedged sample holders with various tilt angles, up to 45$^\circ$ (see Figure \ref{fig:BI:leica}-top). 
The images taken at various incidence angles are shown in Figure \ref{fig:BI:leica}-bottom. In agreement with our simulations described in the previous section, we see that the colour appearance is quite robust for small angles, up to about 20$^\circ$. At already 25$^\circ$, we observe a gradual change in color that is most apparent in the arrays with smaller NW diameter (leftmost arrays). For instance, the middle array (average NW diameter of 130 nm, and 900 nm in pitch) appears bright blue for angles up to 20$^\circ$ and then quickly changes to purple for larger angles of incidence. By contrast, the arrays with larger NW diameters (rightmost) appear more robust in color, where the only the optical density seems to increase with angle. However, as indicated above we suspect the contribution of various NW geometries to the transmission spectrum in these arrays, which may contribute to their robustness in appearance. Introducing multiple NW geometries in a hyperuniform manner to the array may be an interesting additional control knob to design NW-based ST-PV with additional features, such as increased performance or robustness to illumination conditions.

\section{Conclusions}\label{BI:Conclusions}

In this work we explore in detail the potential of semiconductor NW-based solar cells as a powerful and tunable in design for wavelength-selective ST solar cells. The unique property of vertically standing NWs -of diameters in the range of hundreds of nanometers- which is efficient coupling of light into waveguiding modes of these nanostructures makes them great nanophotonics-driven tools to control the light in our intended way. FDTD simulations of PMDS-embedded NW arrays confirm that by changing diameter and periodicity one can engineer the absorption and transmission spectra of a solar cell to provide a broad range of bright colours, semi-transparency and high PV performance.

We find that for three common PV materials (Si, GaAs and InP), when photons with energies close to the material's bandgap efficiently couple to a waveguiding mode of the NW, the photocurrent vs AVT is maximised, and outperforms the idealised non-wavelength-selective performance. In fact, the absorption spectra of the arrays under such conditions resemble those of wavelength-selective approaches, with NIR absorption. From simulations under various illumination conditions (angle and spectrum), we find that the color appearance is quite robust to spectral and small angle changes. For angles larger than $\sim 20^\circ$, the arrays show a more abrupt change in color likely arising from the excitation of Mie resonance modes. Throughout the whole span of illumination angles, the expected photocurrent varies by $\pm 10\%$. 

Finally, we experimentally demonstrate the fabrication and colorful optical properties of free standing vertical GaAs NWs by the PDMS-embedding and peeling-off method. Two sets of purple, blue and green arrays with average AVT around 20 and 50$\%$, respectively, are demonstrated. We compare the transmission spectra and perceived transmitted colours to those expected from geometrical considerations. We also show images of the arrays under increasing angle of incidence up to 45$^\circ$. As predicted by theory, we observe two main regimes: from normal incidence up to 20-30$^\circ$ the colours are kept similar, and for larger angles the color appearance of the arrays changes. Given the fact that high PV efficiencies have already been demonstrated in opaque NW arrays, this work highlights the potential of such nanostructures as new efficient solutions for BIPV. \\

\begin{acknowledgement}

This work is part of the Dutch Research Council (NWO) and was partially performed at the research institute AMOLF.

The authors thank Daniël Koletski for his support with the design and printing of the wedged sample holders, Dr. Lukas Helmbrecht for his help with the optical microscope, and Prof. Albert Polman for fruitful discussions. 
\end{acknowledgement}

\begin{suppinfo}

A listing of the contents of the Supporting Information file is below:
\begin{itemize}
  \item Materials and methods
  \item Photocurrent vs AVT for InP and Si
 \item Chromaticity plots for InP and Si
\end{itemize}

\end{suppinfo}
\newpage

\bibliography{bib2.bib}

\newpage

\section{Supplementary Information}\label{BI:Simulation}

\subsection{Materials and Methods}

\subsubsection {FDTD simulations}
Finite Difference Time Domain (FDTD) simulations of vertically standing nanowire arrays were performed with the Lumerical software package. The simulation geometry consisted of cylindrical nanowires of 6$\mu $m in length, surrounded by PDMS and sandwiched between two layers of indium  tin  oxide (ITO) films of 100 nm in thickness. The nanowires are ordered in a square lattice by using periodic boundary conditions on the x-y boundaries. Perfectly matching layers (PML) are defined at the top and bottom boundaries. The NW diameter is mapped from 50 to 200 nm in steps of 10 nm, and the pitch is changed from 300 to 1000 nm in steps of 50 nm. Each array is excited from the top with a broadband plane wave. The simulation wavelength range was chosen in accordance to the bandgap of each material, where for Si is from 300 to 1100 nm, for GaAs is from 300 to 930 nm, and for InP is from 300 to 1000 nm. The optical constants are taken from Palik (\textit{Handbook of optical constants}). As for direct bandgap materials it is difficult to obtain a good fit to refractive index data due to the sudden jump in the extinction values, the wavelength range for GaAs and InP is divided in two parts to obtain a good fit to experimental data. The mesh accuracy of the simulation layout is chosen large enough for the cylindrical shape of the nanowire to be well-preserved while keeping a reasonable simulation time. Simulations were powered by SurfSara, the Dutch national e-infrastructure with the support of SURF Cooperative.  Frequency-domain field and power monitors are positioned above the source, and before and after the nanowire array to respectively record the reflected power and transmitted power before and after the nanowires. The recorded value shows what percentage of the optical power injected by the source passed through each monitor. By subtracting the absolute values from the two transmission monitors the fraction of absorbed power for each wavelength is obtained. 

\subsubsection {Photocurrent Calculation}

To calculate $J_{ph}$ the fraction of the absorbed photons for each wavelength is obtained from FDTD simulations. By multiplying the absorption with AM1.5G solar spectrum $N_{ph}$ the number of absorbed photons at each wavelength is calculated. After integration over all wavelengths, the total number of generated photocarriers is achieved, and by multiplying by the elementary charge of the electron ($e$) it gives the estimated $J_{ph}$ of our solar cell. 

\begin{equation}
J_{ph} = e \int N_{ph}\left(\lambda\right) abs\left(\lambda\right)  d\left(\lambda\right)
\label{eq:BI:Jsc}
\end{equation}

\subsubsection {Average visible transmittance (AVT) Calculation}\label{BI:AVT}

Average visible transmittance (AVT) is a measure of visible transparency of the semi-transparent solar cell. The recommended calculation approach, which is accepted by window industry is to integrate the transmission spectrum and normalize it to the photopic response of the human eye as such:

\begin{equation}
AVT = \dfrac{\int T\left(\lambda\right) P\left(\lambda\right) S\left(\lambda\right) d\left(\lambda\right)}{\int P\left(\lambda\right) N_{ph}\left(\lambda\right) d\left(\lambda\right)}
\label{eq:BI:AVT}
\end{equation}

\begin{figure}[h]
	\centering
	\includegraphics[width=0.5\linewidth]{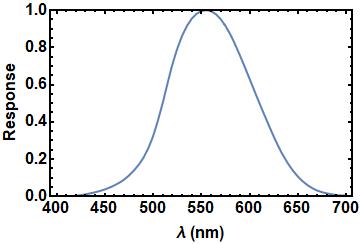}
	\caption{Photopic response of the human eye.}
	\label{fig:BI:photopic}
\end{figure}

Where $\lambda$ is the wavelength of the light, T($\lambda$) is the transmission spectrum (which we obtain from FDTD simulations), N$_{ph}(\lambda)$ is the AM 1.5G solar photon flux and P($\lambda$) is the photopic response of the human eye. The photopic response (sometimes also referred to as y($\lambda$) or V($\lambda$)) is a standard function established by the Commission Internationale de l'Éclairage (CIE) and determines the sensitivity of the human eye to different colors. The CIE photopic standard spectrum of 1978, the most commonly used in window industry, is shown in figure \ref {fig:BI:photopic}. The photopic response is a fairly symmetric gaussian-like distribution centered around 555 nm. 

\subsubsection {Colour Perception Calculation}

\begin{figure}[h]
	\centering
	\includegraphics[width=7 cm]{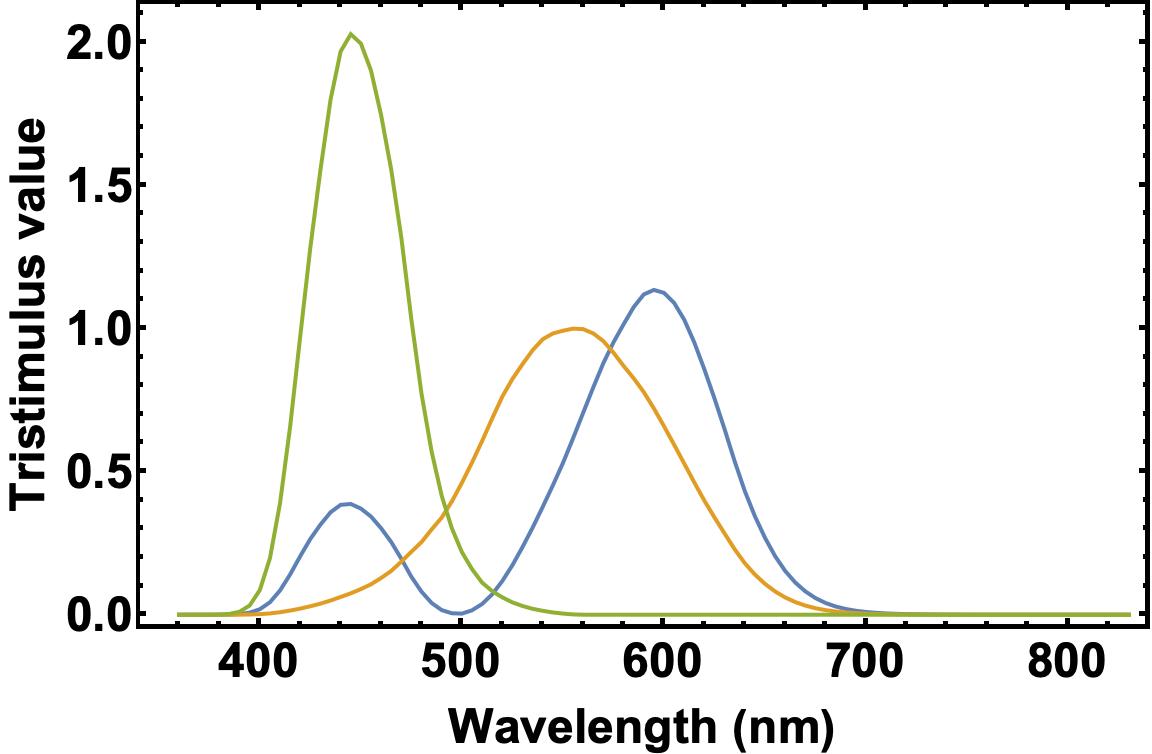}
	\caption{Color matching functions as a function of wavelength}
	\label{fig:BI:tristimuli}
\end{figure}

\begin{figure}[h]
	\centering
	\includegraphics[width=7 cm]{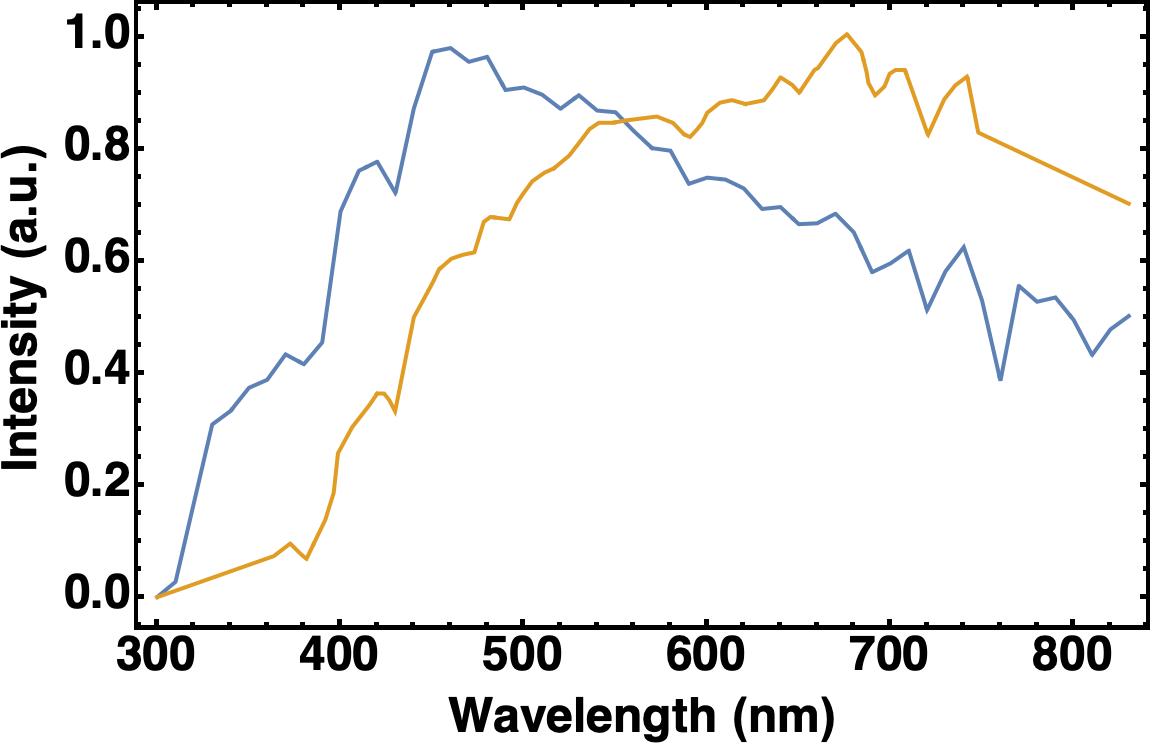}
	\caption{Relative spectral power distribution for standard daylight and evening light illuminations used for our color calculations. CIE standard illuminant D65 is used to represent average daylight.}
	\label{fig:BI:noon}
\end{figure}

The chromaticity coordinates (x,y) are calculated from the spectral power distribution of the stimulus and the CIE color-matching functions (x',y',z') as follows. First, the spectral integration of the color-matching functions convoluted with the stimulus function gives rise the tristimuli values X,Y and Z:
\begin{equation}
\begin{split}
X = \int s(\lambda) x'(\lambda) d(\lambda)\\
Y = \int s(\lambda) y'(\lambda) d(\lambda)\\
Z = \int s(\lambda) z'(\lambda) d(\lambda)
\end{split}
\label{eq:BI:tri}
\end{equation}
The color matching functions are represented in (Figure \ref{fig:BI:tristimuli}). $s(\lambda)$ is the stimulus function, which is basically the light the eye receives. In our case, we consider two different illumination conditions (Figure \ref{fig:BI:noon}) that are normalised by the transmission spectrum of the NW array. The chromaticity coordinates are simply a normalisation of the tristimuli in the form:
\begin{equation}
x = \frac{X}{X+Y+Z}; 
y = \frac{Y}{X+Y+Z}
\label{eq:BI:chrom}
\end{equation}
We have used Mathematica software to obtain the chromaticity values, which are used to represent colored the matrices in Figure \ref{fig:BI:ColourPalette} with the XYZColor function. 

\subsubsection{Nanowire sample fabrication}

\begin{figure}[h]
	\centering
	\includegraphics[width=0.7\linewidth]{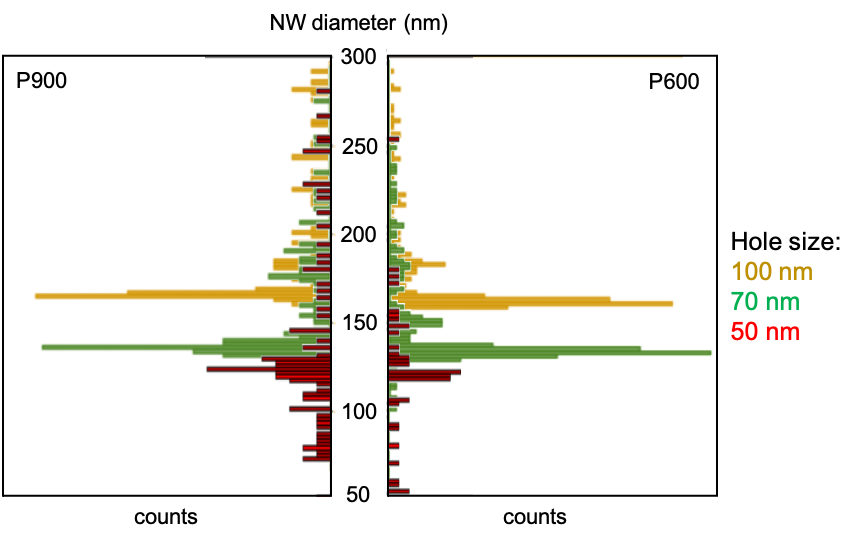}
	\caption{GaAs NW diameter distribution for different hole openings and pitch distances. Values are obtained from image processing of SEM pictures.}
	\label{fig:BI:SI-sizes}
\end{figure}

Arrays of n-GaAs hexagonal pillars were grown by low pressure (20 mbar) MOVPE using trimethylgallium, arsine and disilane as precursor for Ga, As and Si, respectively, in a N$_2$ ambient. Prior to the GaAs:Si growth, the patterned substrates were deoxidized during 4 minutes at 870$^\circ$C under arsine, the decomposition of which results in a H$^+$ surface treatment. The epitaxial growth was subsequently carried out at 850$^\circ$C  at a nominal growth rate of $\sim$46 pm/s (as calibrated separately on a planar, unpatterned GaAs (100) surface).
The resulting sample consists of 9 different growth fields of $300 \times 300$ $\mu m^2$  each with different hole diameters and pitch distances. The NW length is affected by both hole size and pitch distance. For pitch 900 nm, the length of the wires with increasing diameter are respectively around 6.9, 5.7, and 4.5 $\mu m$. The measured NW size distribution is shown in Figure \ref{fig:BI:SI-sizes}.

After growth, the NWs were peeled off from their substrate as follows. First, the as-grown sample was spin-coated with polydimethylsiloxane (PDMS). After spin-coating, the sample was placed in the oven for 24 hours at $50 ^\circ$C followed by at least another 24 hours outside of the oven for complete curing of the PDMS. For the peel off, a small square framing all the NW fields was cut in the PDMS with a surgical knife. Then, the same knife was used to undercut the NW fields from their base and to lift the NWs - PDMS composite from the GaAs substrate. The composite was removed with the help of tweezers and was placed on a clean cover glass slip. If wrinkles are formed or to adjust the position of the composite on the new support, acetone was used.

\subsubsection{Optical characterisation}

A WITec alpha300 RS confocal microscopy setup was used in transmission mode with a 100x and 60x magnification, air objectives, for top and bottom respectively. The sample was illuminated using the integrated LED illumination and spectra were collected using the fiber-connected WITec UHTS spectrometer, where the collection by the fiber acts as the confocal pinhole. The transmission spectra were normalised to the illumination and response function of the set-up by measuring the spectrum without any sample.

For imaging, a Leica DM-LP microscope was used in transmission mode.
An incandescent lamp is used as the light source and the transmitted light is captured by a 20x air objective. The image is captured with Basler Aca1920-40gc camera without any post treatment. For the images taken under different angles, home-made 3D-printed wedged sample supports are used.


\subsection{Chromaticity plots for GaAs and InP}

\begin{figure}[h!]
	\centering
	\includegraphics[width=0.8\linewidth]{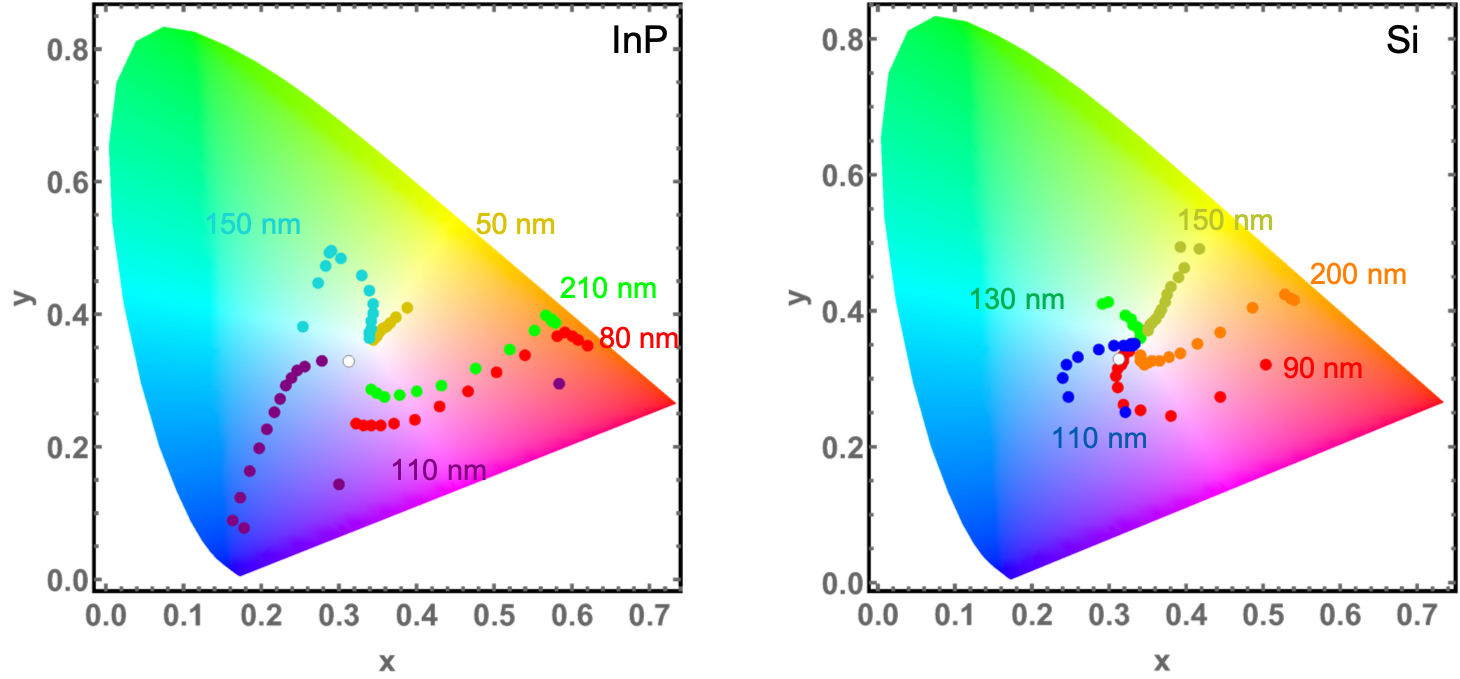}
	\caption{Chromaticity plots for the color appearance in transmission for different NW arrays based on InP (left) and Si (right). The color of datapoints indicates the NW diameter and the pitch distance increases from 300 to 1000 nm from outer-most datapoints towards the center of the chromaticity diagram.}
	\label{fig:BI:SI-chroma}
\end{figure}

The central area of this plot is the area all different white point from different light sources fall into. Thus, the closer to the center the appearance in transmission is, the more transparent. As expected, with increasing the pitch, data points for all diameters move from outside edges of the chromaticity plot towards inside. We observe that for some diameters the hue stays more or less the same but for some other the hue changes quite a bit with the change in pitch. Such changes are softer in the case of Si compared to GaAs and InP, owing to its indirect bandgap nature.

\subsection{Interplay between waveguiding and Mie resonances in the absorption spectra}

\begin{figure}[h!]
	\centering
	\includegraphics[width=1\linewidth]{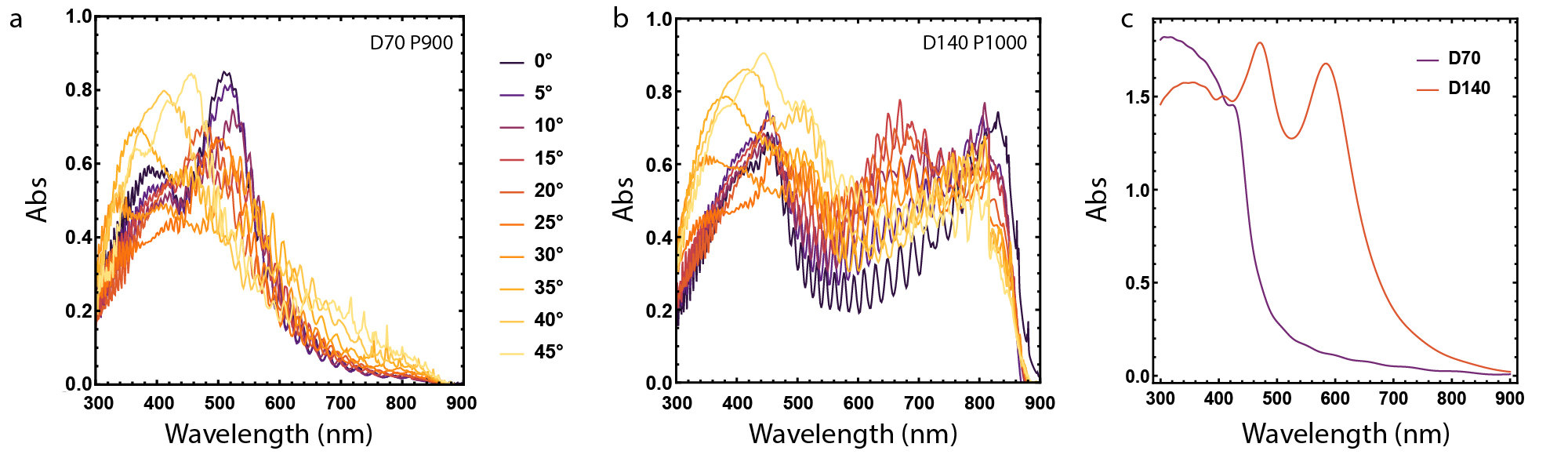}
	\caption{Caption--}
	\label{fig:BI:SI-AngleMie}
\end{figure}

Figures \ref{fig:BI:SI-AngleMie}a and b show the absorption spectra for GaAs NW array with diameter 70 nm and 140 nm, respectively, length of 6 $\mu$m and pitch 900 nm. Different colours are for different angles of incident light. As we see, the peak at ~510 nm which corresponds to light coupling to the first waveguiding mode is very present for the normal incidence. However, by increasing the angle of incidence the light coupling efficiency decreases. From 20 to 25 degrees, however, there is another peak emerging at the lower wavelength range (between 300 nm and 400 nm) in the case of 70 nm NWs. Similarly, for NWs of 140 nm in diameter, additional absorption peaks appear in the green (~500 nm). We assign this new spectral feature to excitation of Mie modes inside the NW. The Mie absorption for GaAs spheres of diameters 70 nm and 140 nm inside PDMS are plotted in figure \ref{fig:BI:SI-AngleMie}c for comparison. The absorption cross-section of the Mie resonators shows similar spectral features to those appearing in the NWs with tilted illumination. 

\subsection{Geometry dependent photocurrent vs AVT for Si and InP}

\begin{figure}[h!]
	\centering
	\includegraphics[width=1\linewidth]{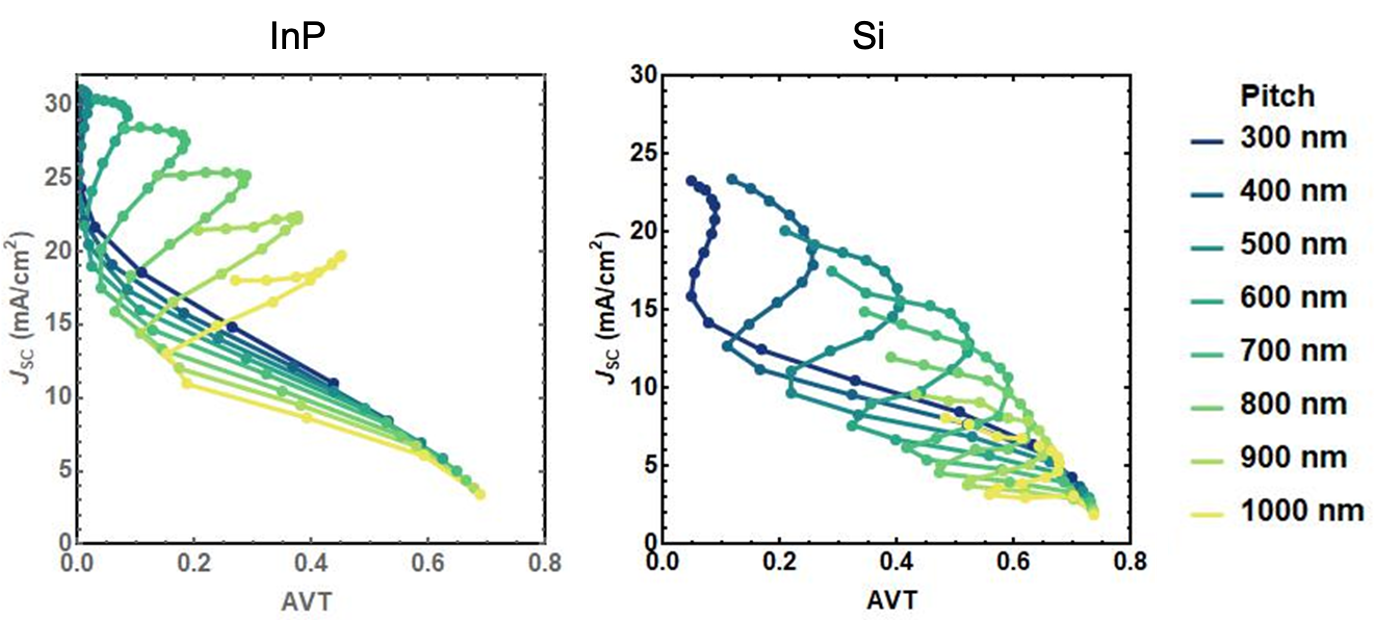}
	\caption{Simulated photocurrent vs AVT for different NW arrays based on InP (left) and Si (right). From blue to green, the color indicates the array pitch distance from 300 to 1000 nm, respectively. The NW diameter is increased from 50 to 200 nm, from right-most datapoints to the left. }
	\label{fig:BI:SI-photoAVT}
\end{figure}

Figure \ref{fig:BI:SI-photoAVT} shows the simulated photocurrent vs AVT for the two other materials: Si and InP. The z-shape described in the main text is less obvious for Si, owing to the indirect nature of its bandgap.

\end{document}